\documentclass[english,prl,reprint,longbibliography,notitlepage,superscriptaddress,preprintnumbers,twocolumn,showkeys]{revtex4-2}

\usepackage{graphicx}
\usepackage{bm}
\usepackage{hyperref}
\usepackage{amsmath}
\usepackage{amssymb}
\usepackage{microtype}
\usepackage{xcolor}
\usepackage[capitalise]{cleveref}
\usepackage{bbm}
\usepackage{mathtools}
\usepackage{dutchcal}
\usepackage{tensor}

\DeclareMathAlphabet{\mathpzc}{OT1}{pzc}{m}{it}

\newcommand{\para}[1]{\par\vspace{2mm}\noindent\textbf{#1}\,---\,}

\makeatletter
\DeclareRobustCommand{\rcite}[1]{%
  \rcite@aux#1,\@nil{#1}%
}
\def\rcite@aux#1,#2\@nil#3{%
  \if\relax#2\relax
    Ref.~\cite{#3}%
  \else
    Refs.~\cite{#3}%
  \fi
}
\makeatother

\hypersetup{
    colorlinks = true,
    citecolor = {red},
    linkcolor = {blue},
    urlcolor = {blue},
}

\begin{document}

\title{
Promise of Future Searches for Cosmic Topology}

\author{Yashar Akrami}
\email{yashar.akrami@csic.es}
\affiliation{CERCA/ISO, Department of Physics, Case Western Reserve University, 10900 Euclid Avenue, Cleveland, Ohio 44106, USA}
\affiliation{Instituto de F\'isica Te\'orica (IFT) UAM-CSIC, C/ Nicol\'as Cabrera 13-15, Campus de Cantoblanco UAM, 28049 Madrid, Spain}
\affiliation{Astrophysics Group \& Imperial Centre for Inference and Cosmology, Department of Physics, Imperial College London, Blackett Laboratory, Prince Consort Road, London SW7 2AZ, United Kingdom}

\author{Stefano Anselmi}
\email{stefano.anselmi@pd.infn.it}
\affiliation{Dipartimento di Fisica e Astronomia ``G. Galilei", Universit\`a degli Studi di Padova, via Marzolo 8, I-35131 Padova, Italy}
\affiliation{INFN, Sezione di Padova, via Marzolo 8, I-35131 Padova, Italy}
\affiliation{LUTH, UMR 8102 CNRS, Observatoire de Paris, PSL Research University, Universit\'e Paris Diderot, 92190 Meudon, France}

\author{Craig J. Copi}
\email{craig.copi@case.edu}
\affiliation{CERCA/ISO, Department of Physics, Case Western Reserve University, 10900 Euclid Avenue, Cleveland, Ohio 44106, USA}

\author{Johannes R. Eskilt}
\email{j.r.eskilt@astro.uio.no}
\affiliation{Institute of Theoretical Astrophysics, University of Oslo, P.O. Box 1029 Blindern, N-0315 Oslo, Norway}
\affiliation{Astrophysics Group \& Imperial Centre for Inference and Cosmology, Department of Physics, Imperial College London, Blackett Laboratory, Prince Consort Road, London SW7 2AZ, United Kingdom}

\author{Andrew H. Jaffe}
\email{a.jaffe@imperial.ac.uk}
\affiliation{Astrophysics Group \& Imperial Centre for Inference and Cosmology, Department of Physics, Imperial College London, Blackett Laboratory, Prince Consort Road, London SW7 2AZ, United Kingdom}

\author{Arthur Kosowsky}
\email{kosowsky@pitt.edu}
\affiliation{Department of Physics and Astronomy, University of Pittsburgh, Pittsburgh, Pennsylvania 15260, USA}

\author{Pip Petersen}
\email{petersenpip@case.edu}
\affiliation{CERCA/ISO, Department of Physics, Case Western Reserve University, 10900 Euclid Avenue, Cleveland, Ohio 44106, USA}

\author{Glenn D. Starkman}
\email{glenn.starkman@case.edu}
\affiliation{CERCA/ISO, Department of Physics, Case Western Reserve University, 10900 Euclid Avenue, Cleveland, Ohio 44106, USA}
\affiliation{Astrophysics Group \& Imperial Centre for Inference and Cosmology, Department of Physics, Imperial College London, Blackett Laboratory, Prince Consort Road, London SW7 2AZ, United Kingdom}

\author{Kevin Gonz\'alez-Quesada}
\affiliation{CERCA/ISO, Department of Physics, Case Western Reserve University, 10900 Euclid Avenue, Cleveland, Ohio 44106, USA}

\author{\"{O}zen\c{c} G\"{u}ng\"{o}r}
\affiliation{CERCA/ISO, Department of Physics, Case Western Reserve University, 10900 Euclid Avenue, Cleveland, Ohio 44106, USA}

\author{Deyan P. Mihaylov}
\affiliation{CERCA/ISO, Department of Physics, Case Western Reserve University, 10900 Euclid Avenue, Cleveland, Ohio 44106, USA}

\author{Samanta Saha}
\affiliation{CERCA/ISO, Department of Physics, Case Western Reserve University, 10900 Euclid Avenue, Cleveland, Ohio 44106, USA}

\author{Andrius Tamosiunas}
\affiliation{CERCA/ISO, Department of Physics, Case Western Reserve University, 10900 Euclid Avenue, Cleveland, Ohio 44106, USA}

\author{Quinn Taylor}
\affiliation{CERCA/ISO, Department of Physics, Case Western Reserve University, 10900 Euclid Avenue, Cleveland, Ohio 44106, USA}

\author{Valeri Vardanyan}
\affiliation{Kavli Institute for the Physics and Mathematics of the Universe (WPI),UTIAS, The University of Tokyo, Chiba 277-8583, Japan}

\collaboration{COMPACT Collaboration}

\date{\today}

\begin{abstract}
The shortest distance around the Universe through us is unlikely to be much larger than the horizon diameter if microwave background anomalies are due to cosmic topology. We show that observational constraints from the lack of matched
temperature circles in the microwave background leave many
possibilities for such topologies. We evaluate the detectability of
microwave background multipole correlations for sample cases.
Searches for topology signatures in observational data over the
large space of possible topologies pose a formidable computational challenge.
\end{abstract}


\preprint{IFT-UAM/CSIC-22-128}

\maketitle

\para{Introduction.}
Standard cosmology combines general relativity and quantum mechanics to produce a simple model accounting for the distribution of matter in the observable Universe. The average spatial curvature of this model is observationally constrained to be flat, or nearly so \cite{Planck:2018vyg}. However, general relativity concerns only the local geometry of the spacetime manifold, not its topology. Quantum processes in the very early Universe may induce ``nontrivial'' (multiply connected) topology of spacetime \cite{HAWKING1978349,Carlip:2022pyh} that remains present today on very large physical scales, even if inflation occurs \cite{Linde:2004nz}. Indeed, the  temperature variations in the cosmic microwave background (CMB) suggest the presence of statistically anisotropic correlations, much as would result from nontrivial topology of comoving spatial sections. These include anomalous statistical properties of  low-multipole harmonic coefficients~\cite{deOliveira-Costa:2003utu,Schwarz:2004gk,Land:2005jq,Planck:2013lks}, lack of large-scale correlations~\cite{Bennett4766,Hinshaw:1996ut,WMAP:2003elm,Copi:2006tu,Copi:2008hw,Copi:2013cya,Planck:2013lks,Planck:2015igc,Planck:2019evm}, and  asymmetry of power on the sky~\cite{Eriksen:2003db,Hansen:2004vq,Park:2003qd,Eriksen:2004df,Eriksen:2004iu,Rath:2007ti,Monteserin:2007fv,Eriksen:2007pc,Hansen:2008ym,Hoftuft:2009rq,Cruz:2010ud,Gruppuso:2013xba,Axelsson:2013mva,Planck:2013lks,Akrami:2014eta,Planck:2015igc,Planck:2019evm,Shaikh:2019dvb,Schwarz:2015cma,Abdalla:2022yfr}. If topology is the explanation for CMB anomalies, there is detectable topological information in the CMB. While unambiguous indicators of topology have yet to be detected, we present evidence that prior searches for topology \cite{Cornish:1997ab,deOliveira-Costa:2003utu,Cornish:2003db,ShapiroKey:2006hm,Mota:2010jb,Bielewicz:2010bh,Bielewicz:2011jz,Vaudrevange:2012da,Aurich:2013fwa,Planck:2013okc,Planck:2015gmu} have far from exhausted the potentially significant possibilities. 
Much more can be done to discover, or constrain, the topology of space.

If the Universe is a manifold with nontrivial spatial topology, then through any spatial point there are closed spacelike curves that are not continuously deformable to a point. 
An observer will perceive each object as having multiple copies, with relative locations determined by details of the manifold. 
This can be interpreted as space having finite extent in one, two, or three dimensions.
The parameter space for such manifolds is much larger than what has been systematically tested. 
Even limited to spatially flat manifolds (i.e., curvature parameter $\Omega_K=0$), there are 17 inequivalent nontrivial topologies, each with multiple real parameters. 
Most attention so far has been confined to the simple three-torus (though see, e.g., \rcite{Aurich:2014sea}) with a rectangular-prism fundamental domain, though a parallelepiped is permitted.

The observed CMB fluctuations encode  information about topology \cite{1993JETPL..57..622S,1993PhRvL..71...20S}, even when the topology scale exceeds the diameter of the visible Universe. CMB observations  probe the last-scattering surface (LSS) of cosmological photons, at a comoving distance of nearly the Hubble scale $H_0^{-1}$ (where $H_0$ is the present value of the Hubble expansion rate). 
Nontrivial topology, by breaking statistical isotropy, induces anisotropic correlations in the CMB temperature and polarization fluctuations. 
When the scale of the topology is small compared to the LSS diameter, pairs of circles with matched temperature (and polarization)  become visible in different parts of the sky~\cite{Cornish:1997ab}. 
These have not been observed \cite{deOliveira-Costa:2003utu,Cornish:2003db,ShapiroKey:2006hm,Mota:2010jb,Bielewicz:2010bh,Bielewicz:2011jz,Vaudrevange:2012da,Aurich:2013fwa,Planck:2013okc,Planck:2015gmu}, but correlations can  persist even when the scale of the topology is large enough to preclude matched circles~\cite{Fabre:2013wia,COMPACT:2023rkp} as might other proposed signatures of topology \cite{2007A&A...463..861R,2023PhRvD.107f3545V}.

In this Letter, we demonstrate that (1) cosmic topology remains eminently detectable despite past negative searches \cite{Cornish:1997ab,deOliveira-Costa:2003utu,Cornish:2003db,ShapiroKey:2006hm,Mota:2010jb,Bielewicz:2010bh,Bielewicz:2011jz,Vaudrevange:2012da,Aurich:2013fwa,Planck:2013okc,Planck:2015gmu}; 
(2) previous analyses have not considered several topological degrees of freedom, even for spatially flat manifolds;
(3) constraints from nonobservation of matched circle pairs are less restrictive than widely believed \cite{COMPACT:2022nsu}; 
(4) anisotropic CMB correlations induced by topology can be  detectable even absent  matched circle pairs \cite{COMPACT:2023rkp,Copi:2024}; and (5) though significantly larger manifolds may be detectable in future observations of large-scale structure \cite{Anselmi:2024}, if topology is the explanation for CMB anomalies the topology scale probably cannot greatly exceed the LSS diameter. Details of the calculations leading to the results of this Letter can be found in \rcite{COMPACT:2022nsu,COMPACT:2023rkp,Copi:2024,Anselmi:2024}.

\para{Cosmic topology.}
Cosmic topology refers to the properties of spatial sections of the $(3+1)$-dimensional manifold describing the Universe on the largest scales. 
We assume the \emph{geometry} of a Friedmann-Lema\^itre-Robertson-Walker (FLRW) metric, with the usual cases of negative, zero (flat or Euclidean), and positive spatial curvature. 
Topologies of these geometries have been widely explored mathematically.
We concentrate here on the flat case, as it suffices to make the  general case for a renewed search for topology.
There are eighteen Euclidean topologies, with up to six real parameters each, and a countable infinity of both spherical and hyperbolic topologies, with just the curvature scale as a parameter \cite{Ellis1971}. 

All possible Euclidean manifolds can be generated starting from 1-2 parallelepiped or hexagonal prisms \cite{Lachieze-Rey:1995,Luminet:1999,Cornish:2003db}. 
In the simplest three-torus ($E_1$),  opposite faces of a right rectangular prism are identified, giving a finite volume and simple periodic boundary conditions.
One can ``tilt'' $E_1$ by starting instead with a parallelepiped. 
Alternately, before identification, one can rotate a rhombic face by $\pi$ (giving $E_2$), a square face by $\pi/2$ ($E_3$), a hexagonal face by $2\pi/3$ ($E_4$) or $\pi/3$ ($E_5$). 
These rotations accompanied by translations are called ``corkscrew'' motions. 
Flipping some faces instead of rotating them generates the Klein spaces ($E_7$--$E_{10}$). 
If one or two dimensions of the parallelepiped have infinite length, space is periodic in the other dimensions; but, one can still rotate or flip the remaining faces ($E_{11}$--$E_{17}$). 
The Hantzsche-Wendt space ($E_6$) uniquely starts from two adjacent rectangular prisms and matches faces between them (see, e.g., \rcite{Aurich:2014sea}). 
In the trivial topology $E_{18}$, also known as ``the covering space,''  any closed  loop can be deformed to a point.

These choices determine various characteristics of the cosmic 3-manifold; it may be orientable (handedness preserving) or nonorientable;
finite in zero, one, two, or three dimensions; 
statistically homogeneous (all observers see the same pattern of their own images around them) or inhomogeneous~\footnote{The Universe can have homogeneous geometry (metric) and stress-energy but still be statistically inhomogeneous.}.

This classification was first applied to cosmology in \rcite{Riazuelo:2003ud}, but certain parameters of some topologies were largely ignored (e.g., the tilt angles of the parallelepiped) as were the consequences of statistical inhomogeneity. These issues will be addressed in detail in \rcite{COMPACT:2023rkp,Copi:2024}.

One can fully describe the topology by determining the images (periodic repetitions) of a coordinate triad based at any point. 
These images need not form a simple periodic lattice. 
We can equivalently relate  topology to tilings of the covering space of the associated geometry.
The individual tiles are called the fundamental domain. 
Despite the name, different shape fundamental domains can tile the covering space yet embody precisely the same set of isometries, and different choices may seem natural to different observers.

For many purposes, we need the eigenmodes of the  Laplacian operator subject to the topological boundary conditions---the analogues of Fourier modes on the covering space \cite{1946PhRv...70..410I}. In a rectangular prism $E_1$, the modes are exactly the Fourier modes, just restricted to a countable set of wave vectors. 
In other compact Euclidean topologies, each eigenmode is a linear combination of a small number of Fourier modes of different equal-magnitude wave vectors. 
In the noncompact topologies $E_{11}$--$E_{17}$, the continuum is modified but not fully discretized. 
These changes to the eigenmodes modify the statistical properties of the matter fields, and are thereby potentially detectable in the CMB.

We concentrate below on $E_1$, $E_2$, and $E_3$ because they suffice to demonstrate our broader points. Each is characterized by  two to six parameters describing the shapes of the faces, and the translations and rotations that carry them between one another. Another three to six parameters may be needed to characterize an observer's position and orientation in the space (see \rcite{COMPACT:2023rkp,Copi:2024} for more details about these and other Euclidean spaces, including the exact definitions of the parameters and the diagrams showing the actions of the transformation generators for the topologies).

\para{Correlations induced by topology.}
For Gaussian fluctuations, two-point correlations contain all the information available to determine the observational repercussions of the topology. Topology breaks the isotropy of covering-space Fourier modes: a continuum of wave vectors becomes a lattice, inducing correlations between amplitudes of spherical harmonics. Even for $E_1$, a parallelepiped fundamental domain skews the grid of allowed wave vectors. Reflections or rotations induce further correlations between Fourier modes, and thus between spherical harmonics, enhancing the prospects for observational signatures in the CMB and large-scale structure.

For the CMB in particular, the correlation of  spherical harmonic amplitudes changes from the diagonal $\left<a_{\ell m}a_{\ell'm'}\right>=C_\ell\delta_{\ell\ell'}\delta_{mm'}$ to a full covariance matrix $C_{\ell m\ell'm'}$. 
Observations indicate that  initial perturbation amplitudes on scales small compared to $H_0^{-1}$ today are well described by a Gaussian distribution following a nearly scale-free spectrum depending only on the amplitude of the wave vector, e.g., as predicted by inflation. With topology, under linear evolution the CMB and large-scale structure would each have an anisotropic Gaussian distribution.

\begin{figure}[hpt!]
    \includegraphics[width=0.5\textwidth]{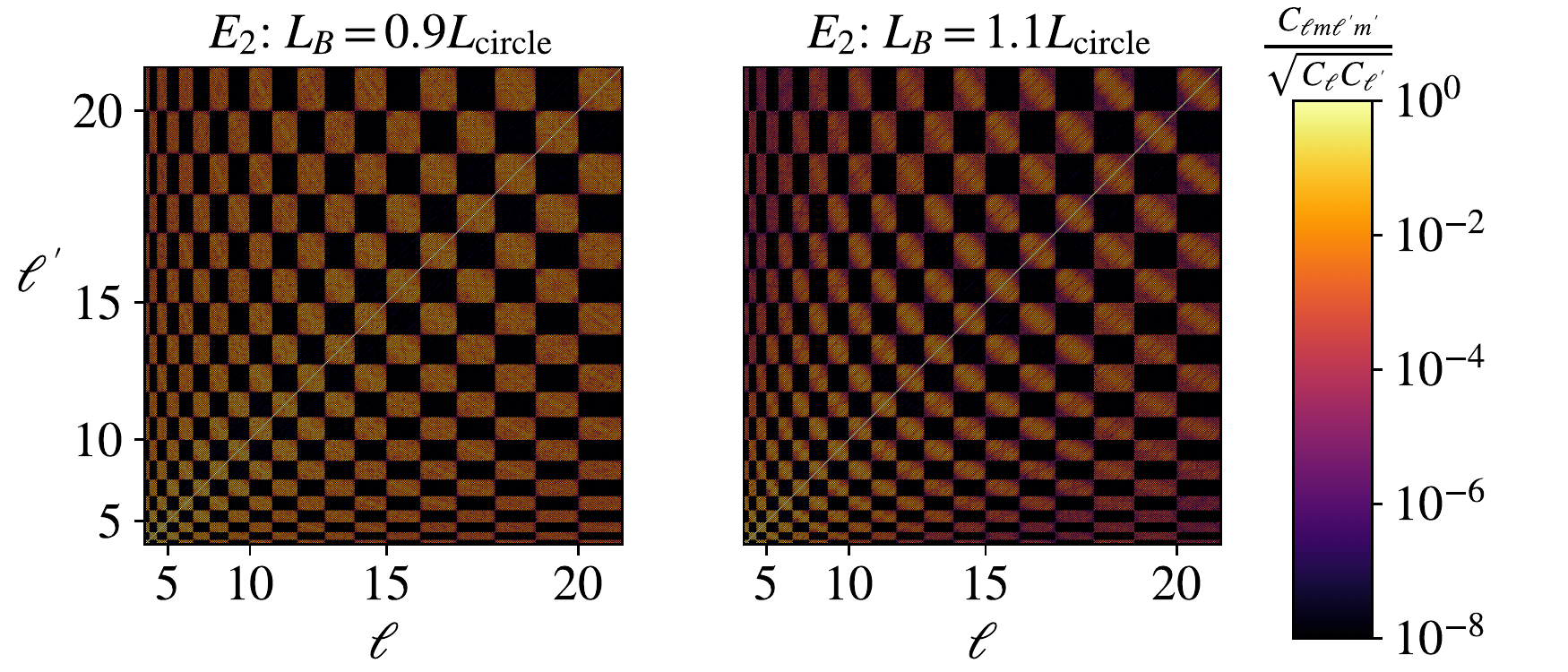}
    \caption{\footnotesize{Portions of rescaled CMB temperature correlation matrices for a half-turn space ($E_2$). 
    $L_{\textrm{circle}}$ is the length scale below which matched circles would be detected;
    $L_B=\{0.9,1.1\}L_{\textrm{circle}}$ is the length along the corkscrew axis;
    $L_A=1.4L_\mathrm{LSS}$ is the other topological length scale.
    The observer is off-axis   at $x_0=(0.35,0,0)L_\mathrm{LSS}$}}
    \label{fig:corr_matrix}
\end{figure}

In \cref{fig:corr_matrix}, we show the rescaled correlation matrices of CMB temperature fluctuations, 
$C_{\ell m\ell'm'}/\sqrt{C_\ell C_{\ell'}}$, induced by (unskewed) $E_2$ topology; see \rcite{COMPACT:2023rkp} for details of calculations.  ($C_\ell$ are elements of the diagonal covariance matrix for the Euclidean covering space computed with the same values of standard cosmological parameters.) 
We consider a right rectangular prism fundamental domain with length $L_A=1.4L_\mathrm{LSS}$  in the two pure translation directions ($\mathbf{\hat{x}}$, $\mathbf{\hat{y}}$), and $L_B=\{0.9,1.1\}L_\mathrm{circle}$ in the corkscrew direction ($\mathbf{\hat{z}}$),
where $L_{\mathrm{circle}}=0.714L_\mathrm{LSS}$ is
the maximum value of $L_B$ for that manifold for which $\geq95\%$ of observers would detect a matched circle pair.
$L_\mathrm{LSS}$ is the diameter of the last-scattering surface of CMB photons.
The observer is placed at $0.35L_\mathrm{LSS}\mathbf{\hat{x}}$ from the corkscrew axis, and does (not) see circles in the left (right) panel. There are significant correlations between disparate $\ell$.

Little work has been done on CMB polarization and topology, but \rcite{Riazuelo:2006tb,Aslanyan:2013lsa}
suggest it provides an additional avenue for exploring topology, and may even have enhanced correlations compared to temperature because polarization originates more predominantly from the LSS\@.

\para{Constraining the topology.}
How can we search for the anisotropic correlations induced by topology? For sufficiently small topology scales, we would see clones of individual objects, such as galaxies or  quasars; these have not been seen \cite{Sokolov:1974,Fang:1983,Fagundes:1987,Lehoucq:1996qe,Roukema:1996cu,Weatherley:2003,Fujii:2011ga,Fujii:2013xsa}. Similarly, in CMB temperature fluctuations, matching pairs of circles on the sky would be signatures of the self-intersection of the LSS\@.
The size of such circles,  the locations of their centers, and the phase of their matching along the circles depend on the details of the topology, but the existence of circles depends only on isotropic geometry and a small-enough topology scale \cite{Cornish:1997ab}. The nonobservation of such matched patterns limits the shortest closed paths across our fundamental domain, which could in turn be interpreted as a limit on the parameters of any given topology.
This search is computationally challenging because of the large number of candidate circle centers, radii, and relative phases, but has been completed on both the Wilkinson Microwave Anisotropy Probe (WMAP) and \textit{Planck} data \cite{deOliveira-Costa:2003utu,Cornish:2003db,ShapiroKey:2006hm,Mota:2010jb,Bielewicz:2010bh,Bielewicz:2011jz,Vaudrevange:2012da,Aurich:2013fwa,Planck:2013okc,Planck:2015gmu}.

Once the length of such a returning path exceeds the diameter of the LSS, there are no longer matching circles, and we must search for excess anisotropic correlation. This can be done by directly evaluating the likelihood as a function of the parameters of the topologies \cite{Phillips:2004nc,Niarchou:2007nn,Planck:2013okc,Planck:2015gmu}. However, this is computationally challenging because we cannot take advantage of the usual simplification of isotropy: the signal covariance matrix is diagonal in harmonic space or, equivalently, it is only a function of the angular distance between pixels. Even calculating and storing the $\mathcal{O}(\ell_\mathrm{max}^4)$ entries of the full $C_{\ell m\ell'm'}$ matrix is infeasible at high $\ell_\mathrm{max}$.

\begin{figure}[t]
    \includegraphics[width=0.45\textwidth]{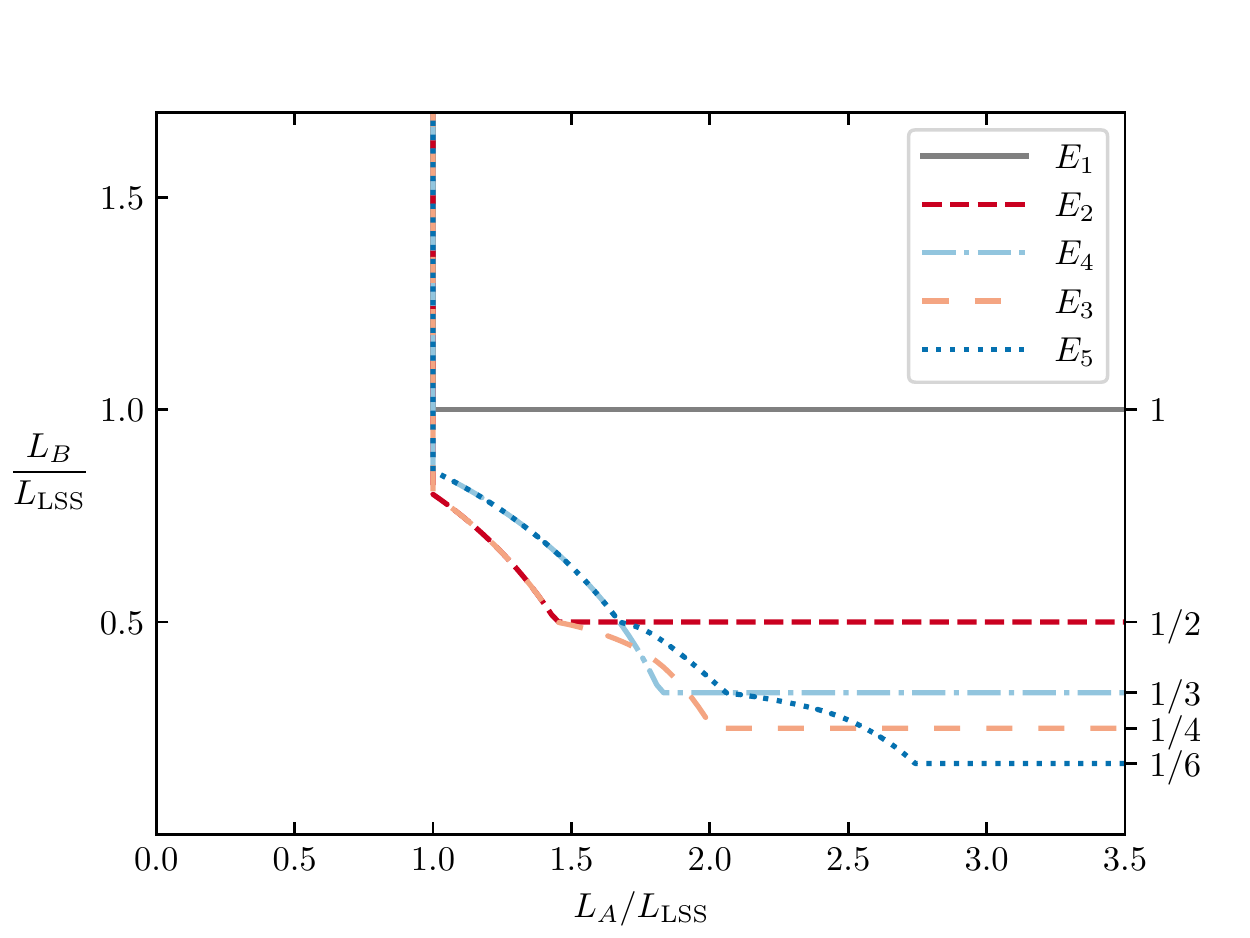}
    \caption{
    Regions of topology parameter space where  observers would or would not see matched circle pairs. For each topology ($E_1$-$E_5$), for values of $L_A$ and $L_B$ in the regions below and to the left of the associated curve, $\geq95\%$ of observers have a clone closer than $L_\mathrm{LSS}$. (In the region to the right of $L_A/L_\mathrm{LSS}=1$ and above $L_B/L_\mathrm{LSS}=1$, no topology will produce any matched circles.)
    For the $E_2$ manifold we choose the right rectangular prism. This figure is identical to Fig. 5 of \rcite{COMPACT:2022nsu}.
    }
    \label{fig:exclusionCurves}
\end{figure}

\para{Current constraints.}
To date, none of the tests outlined above have detected evidence of nontrivial topology.

The matched-circles search has the advantage of being generic: for a sufficiently small fundamental domain, every nontrivial topology of an FLRW cosmology predicts a pattern of repeated circles. However, translating the nondetection of matched circles into limits on topology depends on the details of each individual topology~\cite{COMPACT:2022nsu}. In particular, for topologies in which any of the faces of the fundamental domain is rotated, the induced pattern of circles depends on the location of the observer with respect to the corkscrew  axis---as the observer moves away from the axis, limits on the size of the domain change. In \cref{fig:exclusionCurves}, we display illustrative limits on the parameter space of $E_1$--$E_5$~\cite{COMPACT:2022nsu}: for the excluded parameter values, observers at $\geq95\%$ of locations in the manifold would detect at least one matched circle pair of any size. These curves are indicative of the limits that can be derived from a detailed analysis of matched circles in the CMB temperature from WMAP and \textit{Planck} \cite{Cornish:1997ab,deOliveira-Costa:2003utu,Cornish:2003db,ShapiroKey:2006hm,Mota:2010jb,Bielewicz:2010bh,Bielewicz:2011jz,Vaudrevange:2012da,Aurich:2013fwa,Planck:2013okc,Planck:2015gmu,Petersen:2024b}.
For the simple right-angled $E_1$ this straightforwardly limits the length of the shortest side to the diameter of the LSS sphere, but the rotations in $E_2$ through $E_5$ weaken the limits on the length $L_B$ along the corkscrew axis---as the size of the square or hexagonal face perpendicular to that rotation, $L_A$, increases, observers in more and more of the manifold would not observe matched circles.

\begin{figure}[htpb]
      \includegraphics[width=0.9\linewidth]{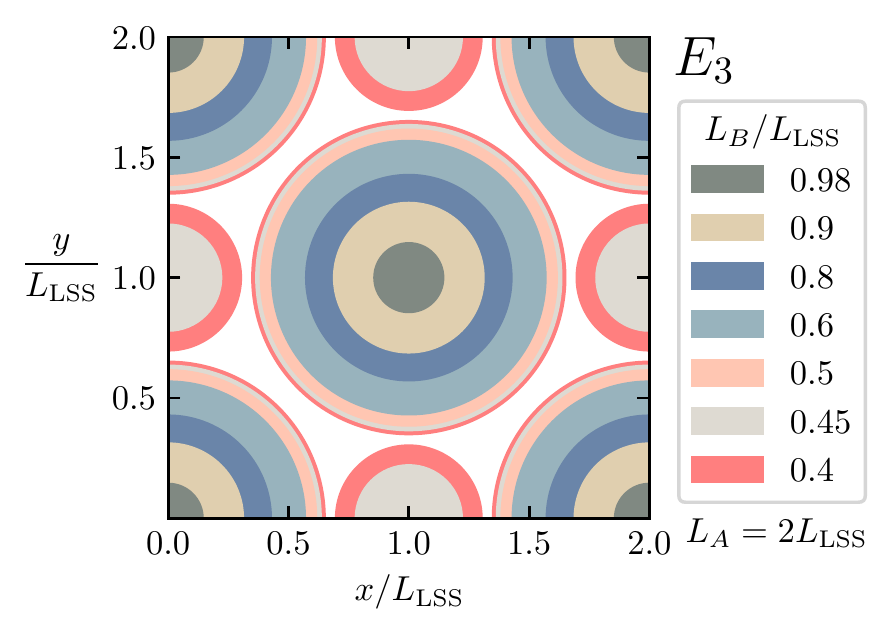}
    \caption{%
    For $E_3$ with $L_A=2 L_{\textrm{LSS}}$,
    locations $(x/L_{\textrm{LSS}},y/L_{\textrm{LSS}})$ of observers in planes of constant $z$ (the corkscrew axis) in which matched circle pairs would be detected, as a function of $L_B$, the length along the corkscrew axis; $L_A$ is the other topological length scale. Even for $L_B=0.4L_{\textrm{LSS}}$ 20\% of observers would still not see matched circles (white regions). This figure is identical to Fig. 3(a) of \rcite{COMPACT:2022nsu}.
    }
\label{fig:figures_E3Contour}
\end{figure}

In \cref{fig:figures_E3Contour} 
we display the square cross section through the $E_3$ manifold with dimensions $L_A=2L_{\textrm{LSS}}$ along both axes, in a plane perpendicular to the axis of rotation at $(x,y)=(L_\textrm{LSS},L_\textrm{LSS})$, where $x,y$ are coordinates in this plane. We shade the regions in which an observer would detect one or more matched circle pairs as a function of $L_B$, the length of the translation associated with the $\pi/2$ rotation.
In regions of a given color observers would see matched circle pairs for all values of $L_B$ less than or equal to the value listed on the legend. As $L_B$ is reduced, more and more observers see matched circles, but even for $L_B=0.4L_{\textrm{LSS}}$ a substantial fraction of the volume is ``circle-free,'' and hence allowed by current observations.

Conversely, the likelihood search is not generic---the covariance matrix must be calculated anew for each topology and each set of topological parameters. Hence, only a very small fraction of testable parameter values \footnote{How small a fraction can only be known once we depart from the special cases that have been studied.} of a subset of Euclidean manifolds have been tested \footnote{Although \rcite{Planck:2015gmu} showed that unrotated slab topologies $E_{16}$ do provide useful proxies for all topologies in which pixel-pixel correlations are predominantly between antipodal regions, and with no change of phase.}.

\para{Future constraints.}
We have already seen in \cref{fig:corr_matrix} that anisotropic correlations persist even when the size of the fundamental domain is larger than the diameter of the LSS and there are no matched circles. Can we detect this richer correlation structure?  The presence of matched circles is a geometric effect, independent of the statistical properties of the fluctuations themselves. Once the scale of the topology is outside of the LSS, we depend on details of statistical properties to detect the induced correlations. We expect this to be more tied to the cosmological parameters than circles. In the following we specialize to models in which the fluctuations are described by the \textit{Planck} 2018 cosmology \cite{Planck:2018vyg}.

We can use the Kullback-Leibler (KL) divergence to compare the probability distribution functions for the $\{a_{\ell m}\}$ in a nontrivial topology, $p(\{a_{\ell m}\})$, and in the trivial topology, $q(\{a_{\ell m}\})$. It is given by 
\begin{equation}
    D_{\mathrm{KL}}(p || q) = \int \mathrm{d}\{a_{\ell m}\} \,\, p(\{a_{\ell m}\}) \ln \left[\frac{p(\{a_{\ell m}\})}{q(\{a_{\ell m}\})} \right]\;,
\end{equation}
the likelihood ratio between the distributions, averaged over $p$ \cite{Fabre:2013wia,Planck:2015gmu}, which describes the information gain from discovering that the Universe is described by a nontrivial topology.
Given data $a_{\ell m}$ from an experiment, $D_{\mathrm{KL}}$ favors nontrivial topology if $D_{\mathrm{KL}} > 1$, giving a threshold for the detectability of nontrivial topology in a perfect experiment with no noise, foreground emission, and  mask. This can be interpreted as a Bayes factor giving the relative odds of the models, or calibrated using Wilks' theorem, which states that twice the likelihood ratio is asymptotically $\chi^2$ distributed with the number of degrees of freedom equal to the difference between those of the two distributions. In our case, this implies a strong detection when $D_{\mathrm{KL}}$ is greater than about three (assuming three degrees of freedom for the orientation, and three more to describe the relevant lengths and angles for the particular topology).

\begin{figure}[hpt!]
    \includegraphics[width=0.5\textwidth]{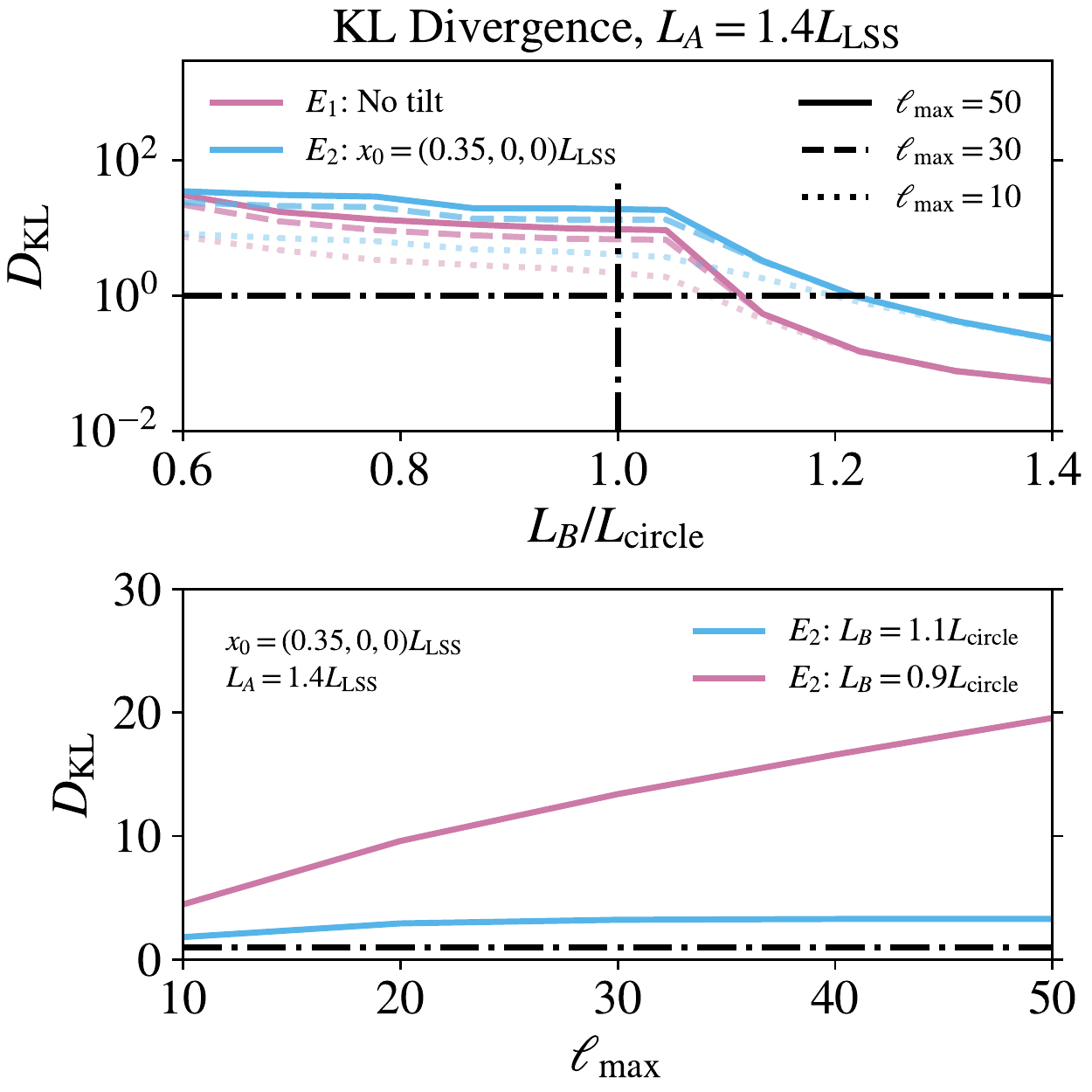}
\caption{\footnotesize{ {\bf Top:} The Kullback-Leibler (KL) divergence as a function of $L_B/L_{\textrm{circle}}$, the length along the corkscrew axis, for $E_2$ and a simple untilted three-torus ($E_1$). $L_{\textrm{circle}}$ is the length scale below which matched circles would be detected. {\bf Bottom:} The KL divergence as a function of $\ell_\mathrm{max}$ for $E_2$.}}
\label{fig:kl}
\end{figure}

\cref{fig:kl} shows the KL divergence as a function of $L/L_{\textrm{circle}}$
and $\ell_{\textrm{max}}$.
They convey a somewhat more optimistic message than  \rcite{Fabre:2013wia} which considered a simple cubic torus $E_1$. We see here (top panel) that $E_2$ domains larger than $E_1$ domains have detectable KL divergence.  
However, we also see that, in $E_2$, once an observer is unable to detect matched circle pairs ($L_B>L_\mathrm{circle}$) only $\ell\lesssim30$ add significantly to the KL divergence. 
Though $D_{\textrm{KL}}$ is challenging to calculate for large $\ell$, 
a signal-to-noise calculation of off-diagonal correlations as signal and statistically isotropic Gaussian random fields as noise gives a similar level of potential detectability \cite{COMPACT:2023rkp}.

The dependence of $D_{\textrm{KL}}$ on  $\ell_\mathrm{max}$ and on topology parameters depends on the specific manifold, and on the observer location within that manifold. 
This is a matter for ongoing investigation, as is the maximum size of each manifold that can be observationally identified.
However, it is important to realize that if topology is the physical cause of CMB large-angle anomalies then ipso facto the CMB must contain detectable topological information.  
Thus the KL divergence would inform us what values of various topological parameters we need to explore for each manifold, avoiding those values for which there is no topological information.
 A topological explanation for any observed CMB anomaly that exhibits parity-violating $\Delta\ell$-odd correlations would only arise from a manifold that encodes parity violation,
 precluding $E_1$, $E_{11}$, $E_{16}$, and, of course, $E_{18}$.
The behavior of $D_{\textrm{KL}}$ with $\ell_\mathrm{max}$ implies that correlations between different $a_{\ell m}$ induced by topology with $L_B>L_\mathrm{circle}$ are limited to low $\ell$, a potential explanation for large-scale CMB anomalies, and holds out the promise for testable predictions with future observations.

Information from the interior of the last-scattering surface should significantly increase  $D_{\textrm{KL}}$, since it will  include many more modes of the same wavelength as those that contributed to the $D_{\textrm{KL}}$ of the CMB. 
In the covering space, we could think of this as resolving the different radial modes that contribute to the same $a_{\ell m}$ of CMB temperature.
In addition, because the projection from the full 3-space to the two-dimensional surface of the LSS mixes many uncorrelated Fourier modes, especially at higher wave number, 
information from shorter wavelength modes will increase $D_{\textrm{KL}}$ by some to-be-quantified amount even for $L_B>L_\mathrm{circle}$.
These questions will be explored in depth in a future publication \cite{Anselmi:2024}.
Future galaxy surveys, 21cm surveys, and other eventual probes of the interior of the LSS thus hold the promise of increasing the range of the topology parameter space that can be explored observationally.

Expanding the reach of past topology searches will be discussed in upcoming work. We anticipate that the dimensionality of the parameter space, the size of the covariance matrix, and its lack of sparsity will make an exhaustive likelihood calculation untenable, even for just Euclidean manifolds. Instead we anticipate developing  machine learning techniques to accelerate the likelihood calculation and to be used in the framework of likelihood-free inference \cite{Alsing:2018eau} to more generally address the challenge of mining CMB data for evidence of nontrivial topology.

\para{Conclusions.}
We have shown that current observations of the CMB have not been comprehensively translated to limits on the allowed large-scale topology of the Universe. We report that for generic Euclidean manifolds, whose isometry groups include rotations or reflections, the lower limit on the topology scale is smaller than the diameter of the LSS by factors of 2--6, and potentially much more. All of the Euclidean manifolds, other than the covering space, violate statistical isotropy; most of them also violate statistical homogeneity. The ability to detect topology even in the absence of explicit matched circles depends on the induced statistical anisotropy. 
The KL divergence suggests that there is information in CMB temperature correlations even when an observer does not see circles, although the topology scale cannot be much larger than $L_\textrm{LSS}$ if topology is to explain CMB anomalies. 
How much this depends on the manifold, its size and other parameters, and the observer's position, is unknown.
Large-scale structure information from future surveys will provide still more information which appears likely to offer a qualitative improvement on CMB temperature correlations. These possibilities will be explored in a series of forthcoming papers.\\

The GitHub repository associated with this study is publicly available at \footnote{\url{https://github.com/CompactCollaboration}}. Codes will be deposited there as publicly usable versions become available.\\

\begin{acknowledgments}
We thank Jeffrey Weeks and David Singer for valuable conversations.
Y.A. acknowledges support by the Richard S. Morrison Fellowship, from research projects PGC2018-094773-B-C32 and PID2021-123012NB-C43, by the Spanish Research Agency (Agencia Estatal de Investigaci\'on)'s Grant No. RYC2020-030193-I/AEI/10.13039/501100011033 and the European Social Fund (Fondo Social Europeo) through the  Ram\'{o}n y Cajal program within the State Plan for Scientific and Technical Research and Innovation (Plan Estatal de Investigaci\'on Cient\'ifica y T\'ecnica y de Innovaci\'on) 2017-2020, and by the Spanish Research Agency through the Grant IFT Centro de Excelencia Severo Ochoa No. CEX2020-001007-S funded by MCIN/AEI/10.13039/501100011033. C.J.C., A.K., and G.D.S.  acknowledge partial support from NASA ATP Grant No. RES240737; G.D.S. from DOE Grant No. DESC0009946; Y.A., P.P., G.D.S., O.G., and S.S. from the Simons Foundation; Y.A., A.H.J., and G.D.S. from the Royal Society (UK); and A.H.J. from Grant No. ST/S000372/1 from STFC in the UK\@.  J.R.E. acknowledges support from the European Research Council under the Horizon 2020 Research and Innovation Programme (Grant Agreement No.~819478). A.T. is supported by the Richard S. Morrison Fellowship. V.V. is supported by the WPI Research Center Initiative, MEXT, Japan and partly by JSPS KAKENHI Grant No. 20K22348. We acknowledge use of the HPC cluster at CWRU\@. 

\end{acknowledgments}

\bibliography{topology,additional}
\end{document}